\documentclass[usenatbib,usegraphicx]{macros/mn2e/mn2e}
\usepackage{amsmath,amssymb,lscape,nth,relsize,algorithm2e,tikz,bm,algorithmicx,algpseudocode}
\title[Fast and reliable symplectic integration]
{Fast and reliable symplectic integration for planetary system $N$-body problems}

\author[David M. Hernandez]
	{David M. Hernandez \thanks{Email: dmhernan@mit.edu} \\ 
	Department of Physics and Kavli Institute for Astrophysics and Space Research,
	\\Massachusetts Institute of Technology, 77 Massachusetts Ave., Cambridge, Massachusetts 02139, USA\\
	}
	\date{Accepted 2016 March 06. Received 2016 February 25; in original form 2015 November 18}

\begin{document}

\maketitle

\label{first page}
\begin{abstract}
 We apply one of the exactly symplectic integrators, that we call HB15, of \cite{HB15}, along with the Kepler problem solver of \cite{WH15}, to solve planetary system $N$-body problems.  We compare the method to Wisdom-Holman methods (WH) in the \texttt{MERCURY} software package, the \texttt{MERCURY} switching integrator, and others and find HB15 to be the most efficient method or tied for the most efficient method in many cases.  Unlike WH, HB15 solved $N$-body problems exhibiting close encounters with small, acceptable error, although frequent encounters slowed the code.  Switching maps like \texttt{MERCURY} change between two methods and are not exactly symplectic.  We carry out careful tests on their properties and suggest they must be used with caution.  We then use different integrators to solve a 3-body problem consisting of a binary planet orbiting a star.  For all tested tolerances and time steps, \texttt{MERCURY} unbinds the binary after 0 to 25 years.  However, in the solutions of HB15, a time-symmetric Hermite code, and a symplectic Yoshida method, the binary remains bound for $>1000$ years.  The methods' solutions are qualitatively different, despite small errors in the first integrals in most cases.  Several checks suggest the qualitative binary behavior of HB15's solution is correct.  The Bulirsch-Stoer and Radau methods in the \texttt{MERCURY} package also unbind the binary before a time of 50 years, suggesting this dynamical error is due to a \texttt{MERCURY} bug.  
\end{abstract}

\begin{keywords}
celestial mechanics - methods: numerical - planets and satellites: dynamical evolution and stability
\end{keywords}

\section{Introduction}
\label{sec:int}
The dynamics of planetary systems, star clusters, or galaxies are gravitational $N$-body problems to first approximation \citep{HH03}.  Gravity determines dynamics on large scales; due to charge screening electromagnetic effects are not relevant at these scales.

Many dynamical problems in science are unified under the Hamiltonian formalism.  A Hamiltonian is a $C^1$ function.  The $N$-body Hamiltonian is
\begin{equation}
H(R,P) = \frac{1}{2} P^{\dag} M^{-1} P + V(R).
\label{eq:hamilt}
\end{equation}
$R$ and $P$ are canonical position and momentum vectors respectively, with size $3N$, $V$ is a potential energy, and $M$ is a positive definite mass matrix of size $3N \times 3N$.  $M$ is diagonal if we work in a coordinate system where the Hamiltonian is quadratic in the momenta; the coordinates which are simplest for our purposes, such as Jacobi coordinates, Democratic Heliocentric coordinates, or inertial Cartesian coordinates, satisfy this criteria.  In inertial Cartesian coordinates, eq. \ref{eq:hamilt} is
\begin{equation}
H(x,p) = \sum_{i=1}^N \frac{p_i^2}{2 m_i} - G \sum_{i=1}^N \sum_{j=i+1}^N \frac{m_i m_j}{\left|x_i - x_j \right|}
\label{eq:hamiltcart}
\end{equation}
For particle $i$, $x_i$ and $p_i$ are its phase space coordinates and $m_i$ its mass.  $G$ is the gravitational constant.  We must solve the coupled system of $6N$ ordinary differential equations resulting from Hamilton's equations: 
\begin{equation}
\begin{aligned}
\frac{d}{dt} R &= M^{-1} P \\
\frac{d}{dt} P &= F(R),
\end{aligned}
\label{eq:ham}
\end{equation}   
where $F(R) = - \nabla_R{V(R)}$.

The solution to equations \ref{eq:ham} cannot generally be written as a quadrature for $N>2$; a practical solution uses numerical integration.  But numerical integration is subject to round-off and truncation error, which in general increase secularly.  Truncation error is due to the finite order of the method and frequently shows linear error growth in energy or angular momentum.  Unbiased rounding error has the same growth in time as Brownian motion; this behavior is also known as Brouwer's law \citep{B37}.  If $t$ is time and $t = hk$ with $h$ a constant time step and $k$ a positive integer, Brouwer's law says that energy, and other first integral, errors grow as $t^{1/2}$ if the truncation error is smaller than the roundoff error.  Biased energy growth is linear in time.  Biased growth is avoided by being careful with the chosen numerical integrator \citep{HMR07, L04}, by using the number format called Universal Numbers, Unum, \citep{G15} or by other methods.     

Geometric integrators \citep{hair06} are the method of choice for solving equations \ref{eq:ham} for long time scales.  One reason is that geometric integrators mitigate secular growth in truncation error.  Rather than focus on small errors which may rapidly accumulate, these maps aim to conserve structure of the governing differential equations.  They address qualitative questions such as ``Will a species survive?'' or ``Is an orbit stable?''  One geometric method is the symplectic method.  Symplectic maps preserve all $n$ Poincar\'{e} invariants, defined over even-dimensional phase space surfaces, including the volume of arbitrary regions of phase space.  $n$ is the number of degrees of freedom: $n = 3N$ in the $N$-body case.  

Symplectic methods have been developed for special astrophysical $N$-body problems.  One example is the time integration of dark matter particles in the cosmological simulation code \texttt{Gadget} \citep{S05}.  \texttt{Gadget} solves the dark matter $N$-body problem where relaxation times are long compared to time scales of interest.  Nearly regular planetary simulations have been successfully studied with the symplectic method of \cite{WH91} (WH), which was further developed by \cite{DLL98} and others.

$N$-body problems with a hierarchy of time scales are difficult to study with symplectic methods.  Changing the time step in effect produces a time dependent effective Hamiltonian \citep{SC93} leading, generally, to a secular growth in the magnitude of energy error.    

Geometric methods are responsible for important solar system dynamics discoveries.  WH was used to show that the orbit of pluto is chaotic \citep{WH91}.  The original WH (WHJ) uses canonical Jacobi coordinates and momenta.  \cite{DLL98} developed a WH version (WHD) by writing the positions as heliocentric coordinates and the center of mass coordinates.  The corresponding canonical momenta are the barycentric momenta and total momentum.  This canonical coordinate system is called Democratic Heliocentric Coordinates (DHCs).  A feature of WH is that the center of mass degrees phase space coordinates do not enter in the equations of motion of the other phase space coordinates so the center of mass phase space coordinates can be ignored.

Although the original WH was pioneering and led to more advanced methods, it has a limited range of usefulness.  WHJ has a greater applicability than WHD.  By changing the numbering of the Jacobi coordinates, WHJ accommodates a small degree of complexity in planetary systems, such as a binary star.  A similar degree of complexity is achieved by the method of \cite{C02}.  But WHJ fails for planetary systems with multiple stars and planetary moons, or systems with non-stellar close encounters within the Hill radii of the objects.  These conditions are common in planetary systems and planetary formation simulations, making WH unsuitable for a wide range of problems.

Switching integrators were developed by \cite{C99} (called \texttt{MERCURY}), \cite{DLL98} (called \texttt{SyMBA}), \cite{K00}, and others to model systems where WH fails.  A switching integrator switches the integration method as a function of phase space coordinates.  For example, \texttt{MERCURY} reduces to WHD when non-solar objects remain well-separated.  Switching integrators are not strictly symplectic because they are not symplectic everywhere in phase space.  The switch can also introduce an artificial drift in error and break time reversibility.  The dynamics in these integrators could be qualitatively wrong.  This is concerning since important dynamics predictions have been derived using switching integrators (\cite{B12}, etc).  We explore these problems with switching integrators in this paper.

Another solution for systems where WH fails is to use general, or hierarchical, Jacobi coordinates, whose structure changes depending on the problem.  Unfortunately, writing a new coordinate system and integrator for each problem is cumbersome.  It would be simpler to have one coordinate system and integrator for all planetary system $N$-body problems.  

\cite{RS15} developed a non-symplectic 15th order integrator and compared its efficiency, the computation time vs the error, with WH.  \texttt{IAS15} is especially efficient for small energy errors (roughly $10^{-10}$ or smaller), as was shown by \citeauthor{RS15}.  To understand the significance of small errors we must first discuss the meaning and usefulness of chaotic $N$-body integrations beyond the integration Lyapunov time.  \cite{QT92} showed that $N$-body orbits are, at least sometimes, shadowed by a true orbit of the original Hamiltonian with close initial conditions to those used in the $N$-body integration.  The smaller the integration errors in the conserved quantities, the closer the integration initial conditions are to the shadow orbit's initial conditions.  More work is needed to determine how shadow orbits apply to more realistic $N$-body integrations.  

Perhaps a more reliable and practical way to understand $N$-body integrations for longer than their Lyapunov time is in an averaged sense; the statistics of $N$-body integrations are assumed to approximate the true statistics of the $N$-body Hamiltonian solution.  The validity of this assumption was recently supported \citep{PB14}.  For integration times longer than the Lyapunov time, a high accuracy solution may not be necessary \citep{PB14}, but errors should remain bounded; symplectic methods are ideal for this requirement.    For some three body problems an energy error magnitude of $10^{-1}$ is sufficient, while for stellar cluster simulations the conventional wisdom is that the energy error magnitude per dynamical time should at most be $10^{-6}$, although this requirement may be too strict \citep{PB14}.  Therefore, a high order code like \texttt{IAS15} may be most useful if both the time scale is shorter than a Lyapunov time and if extreme accuracy is needed, while a lower-order symplectic method may be favored in other cases.  Even under the conditions of high accuracy and short times, a low order symplectic is reliable: we note that the errors of a symplectic method are deceptive.  The errors can often be reduced to the machine precision level by applying a symplectic corrector \citep{WHT96}.  The symplectic corrector only needs to be applied at times when accurate output is needed, after the inverse corrector has been applied at time 0.  Symplectic correctors add negligible computational cost compared with the integration cost. 

\cite{HB15} introduced a symplectic and reversible method (hereafter HB15) for $N$-body problems with small relaxation times.  HB15 was tested on simple regular and chaotic $N$-body problems and the method was found to be efficient against other methods.  Its most straightforward use is with inertial Cartesian coordinates; it is not compatible with Jacobi coordinates or DHCs.  By construction, it exactly conserves all $N$-body integrals of motion, except for the energy, neglecting roundoff and Kepler solver error.  It can be shown that a symplectic method cannot be constructed to conserve all integrals exactly \citep{ZM88}.  HB15 typically requires solving a series of two-body problems with elliptic and hyperbolic orbits.  It was tested as a second order integrator, but it is not difficult to write higher order versions of HB15.  

In this paper we apply HB15 to a variety of problems in planetary system dynamics.  HB15 is optimally used with the Kepler solver from \cite{WH15}, and we do so by default in this paper.  It solves the two-body problem in universal variables without using the Stumpff series.  Other Kepler solvers, such as the one in \cite{dan88} or in \cite{RT15}, fail for some hyperbolic orbits and are not adequate.  The Kepler solver \texttt{drift\_one}, used in \texttt{MERCURY} and a variety of codes, is reliable for hyperbolic orbits, so we use it in a number of tests to see how HB15's performance depends on the Kepler solver.  We use a C version of \texttt{drift\_one}, which is translated, line by line, from the Fortran version.

For problems for which WH methods are designed, HB15 has greater efficiency to a WHD version, and it has the the added benefit of using inertial Cartesian coordinates.  It is not expected to be more efficient than an optimized WH, such as that in \cite{RT15}.  For problems where WH fails, HB15 provides efficiency comparable to or better than \texttt{MERCURY} and several methods for moderate errors.  It is naturally suited for solving problems with a variety of stable hierarchies without having to derive new coordinates and integrators.  It is slower for a problem with frequent close encounters, but errors are small regardless of the closeness of the encounter.  From a practical standpoint, HB15 is simple to implement.  We also compare HB15 against alternatives to switching methods.  We use the efficient two-body solver introduced in \cite{WH15}.  HB15 follows Brouwer's law like \texttt{IAS15}.

The organization of the paper is as follows.  In Section \ref{sec:overview} we provide an overview of HB15 and compare it to WH.  Numerical planetary system comparison tests are carried out in Section \ref{sec:comp}.  Section \ref{sec:switch} discusses switching integrators and their pitfalls.  Section \ref{sec:conc} presents our conclusions and recommendations for numerical methods in planetary system calculations.
\section{Overview of HB15 and comparison with Wisdom-Holman method}
\label{sec:overview}
We compare the steps HB15 and WHD each would take when integrating the Sun and gas giants $N$-body problem. Pseudocode for general HB15 is presented in \cite{HB15} Section 4, and we restate it here.  A drift of time $h$ updates positions by adding a linear term in the momenta.  For HB15, a drift of time $h$ to particle $i$ is $x_i^\prime = x_i + h p_i/m_i$.  The $\prime$ indicates the updated coordinate.  A kick of time $h$ to particle $i$ due to particle $j$ updates $i$'s momentum.  For HB15, $p_i^\prime = p_i + h F_{i j} (x)$ (see eq. \ref{eq:ham}).  The pairs of particles are divided into a kick or Kepler solver group.  We place into the Kepler solver group those pairs that are on a nearly Keplerian orbits or that may have a close encounter.  For a Sun and gas giants integration, the most efficient choice is that the four planet-sun pairs go in the Kepler group, while the other 6 pairs go to the kick group.  The groups cannot change during integration.  Doing so would introduce a time dependence in the integrator ``Hamiltonian'' $\tilde{H}$ \citep{HB15}.    

With these definitions and instructions, we can define a map $\phi_h$, shown in Algorithm \ref{alg:phih}.  We label operators $A_h$, $B_h$, and $C_h$, which will be discussed below eq. \ref{eq:map}.     
\begin{algorithm} 
\caption{Pseudocode for $\phi_h$, the building block of HB15.}
\label{alg:phih}
     \SetAlgoLined
     (Begin $C_h w_x$) \;
     Drift all particles for time $h$  \;
     (Begin $B_h w_x$) \;
     \For{pairs of particles ($k$, $l$) getting kicks}{ 
     		Kick particles $k$ and $l$ for time $h$ using their mutual force \;
     }
     (Begin $A_h w_x$) \;
     \For{rest of pairs ($i$, $j$)}{
 		Drift particles $i$ and $j$ for time $-h$ \;
		Apply a Kepler solver to advance the relative coordinates and velocities of $i$ and $j$ by $h$ \;
		Advance center of mass coordinates of $i$ and $j$ by $h$ assuming their two-body total momentum is conserved \;
		Update inertial Cartesian coordinates and velocities of $i$ and $j$ by using the results of the previous two steps \;
		}
\end{algorithm}
$w_x$ and $w_u$ are the phase space vectors in Cartesian and DHC's respectively.  Transpose map $\phi_h^\dag$ is Algorithm \ref{alg:phih} with the steps reversed.  Then the order 2 method, HB15, is given by   
\begin{equation}
\phi_h^{2 } = \phi_{h/2}^{\dag} \phi_{h/2}.
\label{eq:phi2p}
\end{equation}
$\phi_h^2$ is twice as computationally expensive as $\phi_h$, but it is time reversible.  We find time reversible methods perform better than non-reversible ones.  \cite{hair06} shows reversibility can be as advantageous as symplecticity.  Second order WH methods in theory and practice are also time reversible.  It is unclear how much of their efficiency is due to their symplecticity or reversibility.  

If $A_{h}$, $B_h$, and $C_h$ are operators of time step $h$, both HB15 and WHD are written
\begin{equation}
\phi_h^2 w = C_{h/2} B_{h/2} A_{h/2}^{\dag} A_{h/2} B_{h/2} C_{h/2} w.
\label{eq:map}
\end{equation}
$C_{h/2} w$ is a drift of time $h/2$, which is not equivalent for the two methods.  Let $u$ and $v$ be the position and momentum in DHCs.  Then a drift of time $h/2$ of particle $i$ in WHD gives $u_i^\prime = u_i + h/(2m_i) \sum_{i = 2}^N v_i$.  $i=1$ corresponds to the Sun.  The $C_{h/2} w_x$ drift for HB15 was defined above.  $B_{h/2} w$ is a kick of time $h/2$, which is identical for the two methods, and described above.  The commutator $[B_h, C_h] = 0$ for WHD and WHJ, meaning the order in which the operators are applied does not change the solution, ignoring roundoff error, but generally $[B_h, C_h] \ne 0$ for HB15.    

For WHD there is a simplification in the form of map \ref{eq:map}: $A_{h/2}^{\dag} A_{h/2} = A_h$.  $A_h w_u$ integrates the equations of motion from the Hamiltonian 
\begin{equation}
H(u,v) = \sum_{i=2}^5 \left( \frac{v_i^2}{2 m_i} - \frac{G m_i m_1}{\left|u_i \right|} \right)
\end{equation}
for a time $h$.  These are $4$ independent Kepler problems with 3 degrees of freedom each.  The integration of $A_h w_u$ is done using a Kepler solver such as that described in \cite{WH15}. 

For HB15, writing $A_h w_x$ is more complicated: it is defined in Algorithm \ref{alg:phih}.  The most expensive part for carrying out the resulting steps is solving 4 Kepler problems.  For HB15, $A_{h/2}^{\dag} \ne A_{h/2}$, and twice as many Kepler solver evaluations will be needed, 8 total.  \cite{HB15} found that the time-reversible method $\phi_h^2$ performed better than $\phi_h$, even though they are both second order accurate.

Since $A_h^{\dag} \ne A_h$ for HB15, HB15 uses twice as many Kepler solvers as WH, making HB15 more computationally expensive.  We now explore the computation times of the two methods.  \cite{HB15} found a Kepler solver was $\approx 40 \%$ more computationally expensive than a kick, but this rough figure will vary depending on the Kepler solver, the orbit, the time step, the programming language, and so forth.  With these caveats, we use it to make timing estimates.  A drift had negligible impact on computational resources.  If $n$ pairs are put in the Kepler solver group,  and $t_0$ is the cost of a kick, HB15's cost for one time step is approximately
\begin{equation}
t = \left( 2 \left(\frac{N(N-1)}{2} - n \right) + 1.4 (2 n) \right) t_0.
\label{eq:cost}
\end{equation}
$N = 5$ for the gas giants problem.  WH's cost is less: we replace the $2n$ by $n$.  Therefore, for a Sun and gas giants problem we expect HB15 to be $\approx 42\%$ slower than WHD.  We expect a smaller difference when comparing HB15 with WHJ since WHJ's kick step is more expensive.  Through speed optimizations, we were able to make HB15 faster than WHD in \texttt{MERCURY} in an outer giant planets integration, as shown in Section \ref{sec:outer}.  We do not necessarily expect HB15 to be faster than an optimized WH such as that described recently in \cite{RT15}.   

We describe the speed optimizations.  First is the Kepler solver, described in \cite{WH15}.  A Kepler solver test for elliptical orbits was nearly twice as fast as a \texttt{drift\_one.c}.  For hyperbolic orbits, \texttt{drift\_one.c} was $60\%$ slower.  HB15 is written in C and we used compiler optimizing option \texttt{gcc -O3}.  \texttt{MERCURY} was written in \texttt{Fortran}, which offers in general no speed advantage over C.  \texttt{MERCURY} was run without output in this and all other efficiency tests in this paper to make it as fast as possible.  \texttt{MERCURY} was compiled with \texttt{gfortran}, for this and all other tests in the paper.  Another speed optimization involves the form of eq. \ref{eq:map}.  For the test in Section \ref{sec:outer}, we switched the place of the $C_h$ and $B_h$ operators.  This allows us to combine the kicks form the end of one time step with those of the beginning of the next time step, so that the total number of kicks is essentially cut by half.  This leads to a new eq. \ref{eq:cost} that is approximately
\begin{equation}
t = \left( \frac{N(N-1)}{2} - n + 1.4 (2 n) \right) t_0.
\label{eq:cost2}
\end{equation}

As the number of Solar System planets increases, the timing difference between HB15 and WH, according to eq. \ref{eq:cost} becomes smaller as kicks dominate computation time.  Using arguments above, kicks are responsible for more compute time than the Kepler solvers for $N > 5$ for both methods.  As $N$ increases, the relative kick contribution also increases.  A benefit of HB15 is that it restores the problem's center of mass degrees of freedom and conserves them exactly, making it easy to follow motion in arbitrary inertial frames.    

Ignoring roundoff and Kepler solver error, HB15 solves the two body problem exactly.  As pointed out by \cite{W06} and as we check analytically, WHD does not.  We can test this using \texttt{MERCURY}'s WHD.  We integrate a Jupiter mass planet at 1 AU in circular orbit around the Sun for one time step $h = 10$ years and we get $|\Delta E/E| = 8.9 \times 10^{-11}$, more than five orders of magnitude larger than machine precision.  HB15's solution that results from integrating the same problem for the same $h$ has $|\Delta E/E| = 1.1 \times 10^{-16}$, the machine precision.

It will be useful for later to discuss the conservation of angular momentum when applying WHJ.  As shown in \cite{HB15}, HB15 conserves angular momentum exactly.  WHJ has analogous operators $A_h$, $B_h$, and $C_h$ of eq. \ref{eq:map} \citep{RT15} corresponding to exact integration of time $h$ of the equations of motion of three Hamiltonians, $H_A$, $H_B$, and $H_C$.  Let $a$ and $b$ be a canonical set of Jacobi coordinates.  With $i$ a particle number, the angular momentum vector satisfies $L = \sum_i x_i \times p_i = \sum_i a_i \times b_i$.  Then it is straightforward to show that $\{L,H_A\} = \{L,H_B\} = \{L,H_C\} = 0.$  The implication is that WHJ conserves angular momentum exactly.  Symplectic correctors, also composed of $A_h$, $B_h$, and $C_h$ operators, do not affect the angular momentum conservation.    
\section{Comparison Tests of Methods} 
\label{sec:comp}  

\subsection{Hierarchical Triples}
\label{sec:hier}
Most planetary systems probably have at least one hierarchical triple.  One example is a planet orbiting a binary star.  A planet-moon configuration orbiting a star is another hierarchical triple.

These systems are quasi-periodic, meaning the Kolmogorov-Arnold-Moser (KAM) theorem can apply.  The KAM theorem says that if a sufficiently differentiable map is perturbed from an integrable problem, invariant curves persist if the perturbation is small and the rotation number is not close to a rational number.  A symplectic map is especially suited for such a problem in long term dynamics over a number of orbits where a standard integrator's accumulated truncation error is too large.  First we study the hierarchical triple system with parameters described in \cite{DLL98}, which is an artificial test problem.  A binary planet system initially has semi-major axis $a = 0.0125$ AU and eccentricity $e = 0.6$.  It is at apocenter, aligned along the $x$-axis, with a counter-clockwise motion in the $x-y$ plane.  Its center of mass is initially in a counterclockwise circular orbit also in the $x-y$ plane with the Sun.  The Sun is at 1 AU from the binary planet center of mass.  The planets have Jupiter masses.  

\begin{figure}
 \includegraphics[width=.5\textwidth]{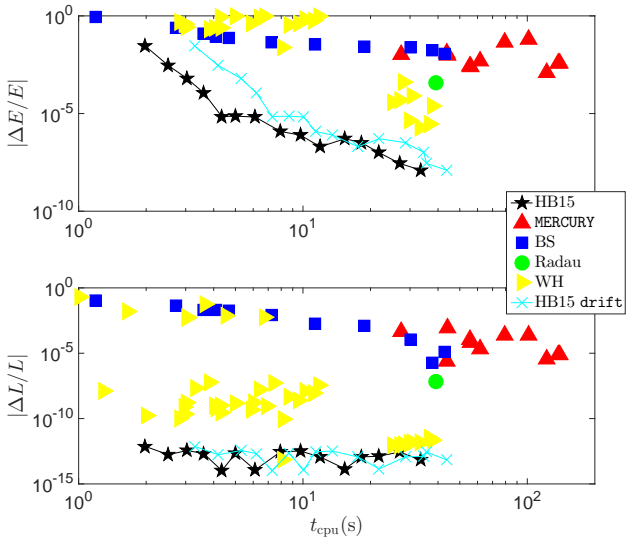}
  \caption{Efficiency for several planetary system integrators when solving the binary planet hierarchical triple system problem.  The problem is run for 10,000 years.  HB15 is most efficient in the range plotted.}
  \label{fig:hier}
  \centering
\end{figure}

We apply different methods to evolve this system: HB15, \texttt{MERCURY} \citep{C99}, and other methods in the \texttt{MERCURY} software package.  The other methods are Bulirsch-Stoer (BS), Radau, and WHJ with symplectic correctors to third order and a default Jacobi numbering.  The default numbering gives the Sun the first coordinate.  Although we do not use \texttt{IAS15} we do use a Radau method.  \texttt{IAS15} is based on the Radau method of \cite{E85}.  The \texttt{MERCURY} switching integrator is frequently used to solve a variety of planetary system problems in the literature.  We also use HB15 with \texttt{drift\_one}.  The three pairs of particles are placed in the Kepler solver group in HB15.  One pair describes elliptic motion while we found the other two always describe hyperbolic motion.  We run the problem for 10,000 years, which corresponds to about 320,000 orbits of the binary.  We vary the time step or tolerances in the methods and measure the efficiency in calculating the energy and angular momentum.  The errors are output by the \texttt{MERCURY} software and we copy them directly.  Results are shown in Fig. \ref{fig:hier}.

All methods give the qualitatively correct answer that the tight binary remains bound.  The slope of the least squares linear fit of the HB15 points is $-4.5$: it is behaving better than a second order method for this problem.  HB15 is the most computationally efficient method for the given range of energy errors.  HB15 with $\texttt{drift\_one}$ is 30 - 70$\%$ slower, but still is preferable against the other methods.  At smaller errors, HB15 loses its efficiency, as expected, since it is a low order integrator.  A symplectic corrector could reduce the HB15 errors by orders magnitude at virtually no additional computation cost.  We leave development of symplectic correctors for HB15 for future work.  The WHJ points are divided into two clusters.  The first cluster points correspond to step sizes that work out to $\ge 36$ steps per binary planet period, while the second cluster points with the larger error correspond to $\le 34$ steps per binary period.  Following \cite{W15}, we can calculate an effective pericenter period, $T_{\dot{f}}$.  The 35 steps per period are 7 steps per $T_{\dot{f}}$, so that 7 steps per $T_{\dot{f}}$ are required to begin resolving pericenter.  The critical radius for \texttt{MERCURY} \citep{C99}  is set at the default $r_{\mathrm{crit}} = 3 r_{\mathrm{Hill}}$, where $r_{\mathrm{Hill}}$ is the Hill radius.

HB15 usually gives the smallest angular momentum errors, by orders magnitude.  We note that the error in angular momentum for WHJ (labeled as WH) is a bit large given our predictions in Section \ref{sec:overview} that neglecting roundoff and Kepler solver error, the angular momentum error should be 0.  This large error was also observed in \cite{RS15}, Fig. 4, although they did not point out it was unexpected.  We posit the WHJ implementation of \texttt{MERCURY} may have a bug.    

\begin{figure}
 \includegraphics[width=.5\textwidth]{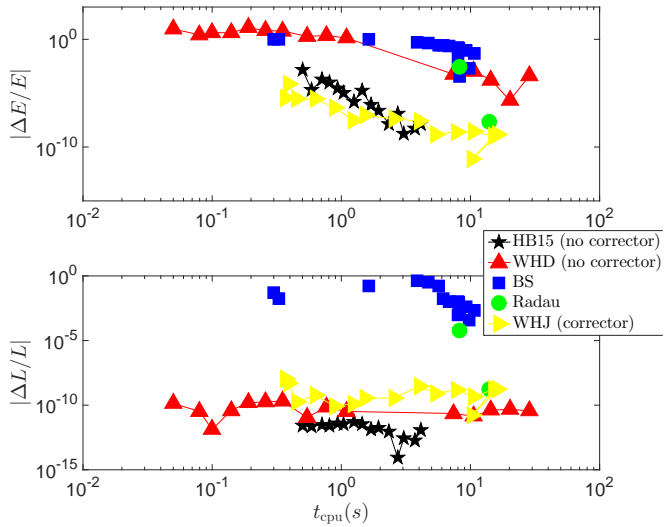}
  \caption{Efficiency, as in Fig. \ref{fig:hier} for the solution of a hierarchical triple problem consisting of a binary star orbited by a planet.  The problem is run for 100,000 years.  HB15 and WHJ, with symplectic correctors, are best suited for this problem in the range plotted.}
  \label{fig:binstar}
  \centering
\end{figure}

We see qualitatively similar results when we test a binary star orbited by a planet.  We let the binary stars have $a = 1$ AU and orbit each other with $e = 0.6$.  They start at apocenter aligned with the $x$-axis, and orbit counterclockwise in the $x-y$ plane.  A planet is set at $a = 20$ AU in the $x$ direction from the stars' center of mass, $C$.  The planet-$C$ system also has counterclockwise motion in the $x-y$ plane with $e = 0.6$.  The stars have solar masses while the planet has Jupiter mass. We run the problem for 100,000 years, which is 100,000 orbital periods of the binary: a similar order magnitude to the number of orbital periods of the binary planet of the previous problem.  The results are shown in Fig. \ref{fig:binstar}.  In this case, WHJ is approximately as efficient as HB15.  However, its slope is $-2.7$, while HB15 again acts like a higher order integrator, with slope $-5.8$, meaning it becomes comparatively better as $h$ is reduced.  Note that WHJ, unlike HB15, has symplectic correctors applied which reduce its errors.  The \texttt{MERCURY} switching integrator run with $r_{\mathrm{crit}} = 3 r_{\mathrm{Hill}}$ reduces for this problem to WHD.  We checked the close encounter flag is never activated.  It is less efficient than HB15.  Because this WHD has had no symplectic correctors applied to it, like HB15, it may be a fairer comparison to HB15 than WHJ with correctors.  HB15 has the lowest angular momentum errors.  The next smallest errors are given by WHD and WHJ.  

\subsection{The Outer Giant Planets}
\label{sec:outer}
We began discussion of the problem of the Sun and outer gas giant planets in Section \ref{sec:overview}.  WH uses 4 Kepler solvers per step while HB15 uses 8.  We place the planet-sun pairs in the Kepler solver group to obtain the most efficient solutions.  As described in Section \ref{sec:overview}, we optimize HB15 by switching the order in which we apply the $B_h$ and $C_h$ operators, essentially reducing the number of kicks by half.

\begin{figure}
 \includegraphics[width=.5\textwidth]{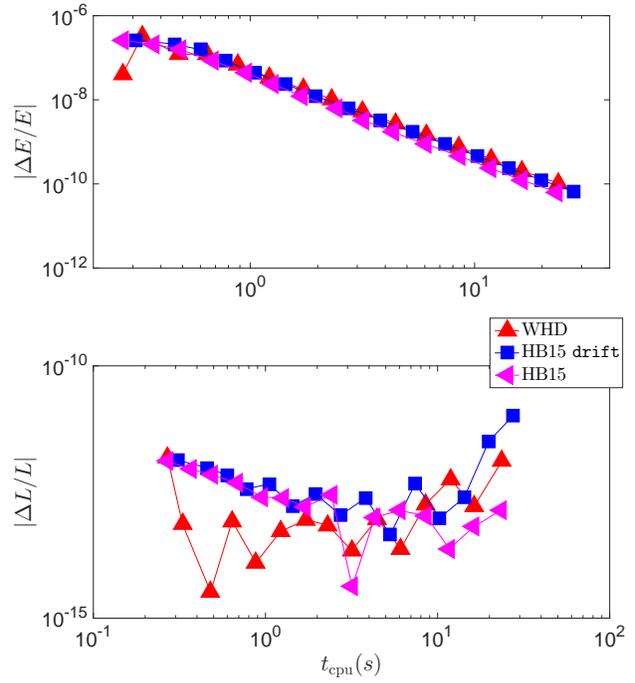}
  \caption{Efficiency plot for the the solution of outer giant planet $N$-body problem.  The problem is run for 100,000 years.  HB15 and WHD have similar efficiency.}
  \label{fig:outerp}
  \centering
\end{figure}

We integrate the outer gas giants for 100,000 years for different time steps.  To use WHD, we ran \texttt{MERCURY} with its $r_{\mathrm{crit}}$ parameter set to  $0$.  Initial conditions were taken from \cite{hair06}.  Also, we again run HB15 with $\texttt{drift\_one.c}$ for reference.  The efficiencies of the solutions are shown in Fig. \ref{fig:outerp}. 
The slopes of the three curves in the energy plot are $\sim -2$, which is what we expect for second order methods.  The efficiency of the two methods is similar.  The slope of a least squares fit line to the points in the energy plot is $-2$, which is what we expect for second order methods.  HB15 is the fastest method while HB15 with \texttt{drift\_one} is approximately equally fast or marginally faster than WHD.

\begin{figure}
 \includegraphics[width=.5\textwidth]{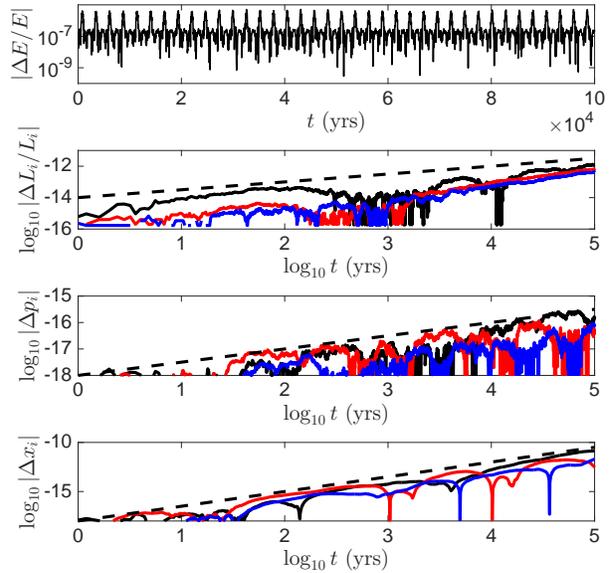}
  \caption{Error in conserved quantities as a function of time for HB15 applied to the outer giants problem.  $h = 0.1$ years.  We see the expected bounded energy error and machine precision error in other conserved quantities (ignoring roundoff error).}
  \label{fig:outer4}
  \centering
\end{figure}

HB15 should show non-secular growth in energy and machine precision accuracy in the other integrals of motion neglecting roundoff and Kepler solver error.  We check this by calculating the error in conserved quantities as a function of time when $h = 0.4$ years.  The result is shown in Fig. \ref{fig:outer4}.  The top panel shows the energy error, calculated every 100 steps to show structure of the error curve. This panel can be compared to Figure 2 of \cite{DLL98}.  HB15's median $\Delta E/E$ is about $10^{-7}$ while WHJ (without correctors) and WHD's median appears to be $2 \times 10^{-7}$ and $0.5 \times 10^{-7}$ respectively in that paper.  The $L_i$, $p_i$, and $x_i$ are the three components of angular momentum, linear momentum, and center of mass position constant, respectively.  In this Section, we have shifted to an inertial frame where the initial $p_i$ and $x_i$ are 0, although the shift is not required for HB15 to work well. We see the expected behavior of symplectic and reversible methods.  The negative dips in the bottom three panels indicate changes of sign.  The growth in error in the integrals besides energy is due to roundoff error.  Before we examine the growth of roundoff error in Fig. \ref{fig:outer4}, a discussion on biased and unbiased roundoff error growth is in order.   

Tests in \cite{HMR07} showed symplectic compositional methods follow Brouwer's law.  It remains to be seen whether HB15, also a symplectic compositional method, has the same behavior.  Brouwer's law means the growth in energy error is a random walk with mean zero and variance proportional to the roundoff unit.  For symplectic methods following Brouwer's law, random walk behavior is achieved for other conserved quantities of the symplectic method besides the energy if they are near the conserved quantities of the continuous problem.  Such conserved quantities in HB15 are angular momentum and linear momentum.  In Fig. \ref{fig:outer4}, the growth in time of linear momentum is $1/2$, as expected for a random walk.  The dashed line indicates slope $1/2$.  The center of mass position is expected to grow as $t^{3/2}$ because it is an integral of the linear momenta over time.  In agreement with this expectation, the center of mass position error in the 4th panel of Fig. \ref{fig:outer4} grows as $t^{3/2}$: the dashed line indicates a slope $3/2$.  The story for the angular momentum error is different: the dashed line indicates slope $1/2$, and matches well the initial behavior.  After $\log_{10} \left(t/\mathrm{yrs} \right) = 3$ the slope of the error becomes steeper, although still less than linear which is the largest theoretically possible roundoff growth rate \citep{H62}.  One source of the bias is the Kepler solver.  If we simply solve a two-body problem with $e = 0.2$, $GM = 1$, $a=1$, $h = P/10$, and $t_{\mathrm{max}} = 10^5 P$, with $M$ the total mass and $P$ the orbital period, the energy and angular momentum grow with time due to roundoff with slope between $0.5$ and $1$.  

\begin{figure}
 \includegraphics[width=.5\textwidth]{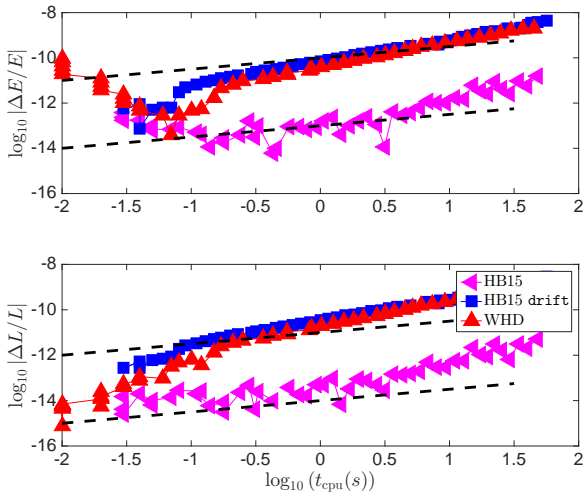}
  \caption{Efficiency plot for the the solution of outer giant planet $N$-body problem.  The problem is run for 10 years with small step sizes.}
  \label{fig:roundoff}
  \centering
\end{figure}

A further roundoff error test comes from redoing Fig. \ref{fig:outerp} with smaller $t_{\mathrm{max}}$, $t_{\mathrm{max}} = 10$ years, and smaller step sizes such that roundoff error has a greater error impact.  Fig. \ref{fig:roundoff} shows the results.  Roundoff error is a function of the number of steps $n$.  In the case of the methods of Fig. \ref{fig:outerp}, $n$ is proportional to $t_{\mathrm{max}}$, which is approximately proportional to the cpu time.  
So the dashed line, which has slope $1/2$, approximately indicates Brouwer's law.  HB15 with \texttt{universal.c} has errors that are about 3 orders of magnitude smaller than the other methods for a given $t_{\mathrm{cpu}}$, but all methods are biased.  Roundoff error in energy is noticeable if it is at the level of truncation error, which is typical of a high order method like \texttt{IAS15}, and will not impact as often lower order integrators like HB15, which are better suited for qualitative studies.

The error in energy in the top panel of Fig. \ref{fig:outer4} is bounded and remains bounded for larger $h$.  HB15 should be symplectic and show favorable long term energy conservation regardless of what pairs are put in the Kepler group in Algorithm \ref{alg:phih}.  If we put all pairs in the solver group, HB15 is unnecessarily slow.  With all pairs in the Kepler solver group of Algorithm \ref{alg:phih}, we choose $h = 0.1$ years and after running HB15 find expected error behavior similar to that in the top panel of Fig. \ref{fig:outer4} (not shown here).    
\subsection{The Outer Gas Giants with a Binary Planet}
We increase the complexity of our tests by combining the binary planet problem from Section \ref{sec:hier} and the outer giant planets problem from Section \ref{sec:outer}; the particles are the Sun and 6 total planets.  The same pairs are left in the Kepler solver group: there are 21 pairs, 8 solved with Kepler solvers.  WH gave large error for all practical time steps we tried for this problem and we must use \texttt{MERCURY} or a similar method for comparison.  We set $r_{\mathrm{crit}}$ at the default $r_{\mathrm{crit}} = 3 r_{\mathrm{Hill}}$.

\begin{figure}
 \includegraphics[width=.5\textwidth]{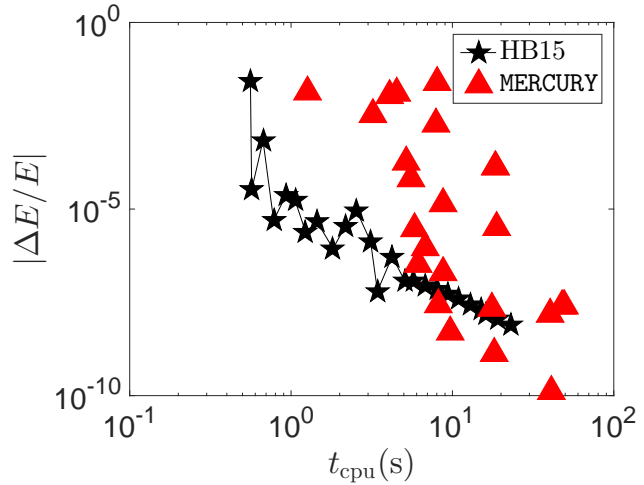}
  \caption{Efficiency of \texttt{MERCURY} and HB15 for the solution of the outer gas giants with a binary planet problem.  The problem is run for 1000 years.  HB15 is most efficient at larger error ranges.}
  \label{fig:outerplusbin}
  \centering
\end{figure}

We tried other complex problems such as the outer gas giants plus Pluto and Charon or the outer gas giants plus the Earth and Moon.  The qualitative efficiency results were not different.  We run the problem for 1000 years, which is $\approx 23,000$ orbits of the binary.  Efficiency plots are shown in Fig. \ref{fig:outerplusbin}.  

The \texttt{MERCURY} points are calculated by varying its $h$ and Tol simultaneously.  The most efficient \texttt{MERCURY} points have the smallest Tol tested, Tol = $10^{-12}$.  The slope of the HB15 curve is $-2.2$, consistent with a second order method.  Fitting a line to the \texttt{MERCURY} points gives an unreliable $-4.2$ slope.

\subsection{The Massive Outer Gas Giants}
We tested a massive outer solar system, where the planet masses are increased by a factor 50.  This problem was studied in \cite{DLL98} and \cite{C99} and is a good test to check how the code handles close encounters between non-stellar objects.  The problem solution is chaotic and it is not useful to look at one set of initial conditions.  The chaos is due to the numerical method and the physical continuous Hamiltonian.  Because of numerical chaos that depends on the method, it is difficult to compare methods for this problem.  

To measure chaos, we calculate the growth in the $L_1$ norm, which is a measure of phase space distance.  A precise definition, although the exact normalizations will not matter for our purposes, is
\begin{equation}
L_1(t) = \sum_{j = 1}^{N_{\mathrm{dim}}} \sum_{i = 1}^{N} \left( \frac{\left|x_{i j}(t) - x^\prime_{i j}(t) \right|}{\left|x_{i j}(0) - x^\prime_{i j}(0) \right|} + \frac{\left|p_{i j}(t) - p^\prime_{i j}(t) \right|}{\left|p_{i j}(0) - p^\prime_{i j}(0) \right|} \right).
\end{equation}
The $\prime$ indicates a perturbed trajectory.  $i$ is the particle number, $j$ is the dimension, and $N_{\mathrm{dim}}$ is the number of dimensions in the problem, typically 3.  If necessary one reinitializes $L_1(t)$ when it is too large.  Initially, the perturbed trajectory differs in one phase space variable, say by $10^{-14}$, and we check that the behavior of $L_1(t)$ is consistent when perturbing other phase space variables.  \cite{chan90} argue that symplectic integrators are preferred over non-symplectic integrators for chaotic problems.  For $h = 0.001$ years, during the first 100 years, the growth in the norm was between linear and quadratic, a growth characteristic of regular problems.  For $h = 0.005$ years, during the same time period, the growth in the $L_1$ norm was exponential with Lyapunov time scale of about $40$ years, consistent with chaos.  

We set 30 different initial conditions for the massive outer gas giants problem.  They are obtained by perturbing each of the 30 positions and velocities by $1 \times 10^{-14}$.  We run the problem for 3000 years, as is done in \cite{DLL98}.  We use a step $h = 0.005$ years, which led to chaotic behavior above.  We also tested larger and smaller time steps, which yielded larger and smaller energy errors, respectively.  The total number of steps is $6 \times 10^5$.  We note that a non-symplectic adaptive time step method could be an adequate method for solving this problem if its total number of time steps is also $6 \times 10^5$.  The method's suitability and efficiency would have to be tested by numerical experiment.  Close encounters can occur between any pairs so we place all pairs in the Kepler solver group of Algorithm \ref{alg:phih}.  With all pairs in the Kepler solver group, HB15 will compute at its slowest (see eq. \ref{eq:cost} with $n = N(N-1)/2$, $n$'s maximum value).

\begin{figure}
 \includegraphics[width=.5\textwidth]{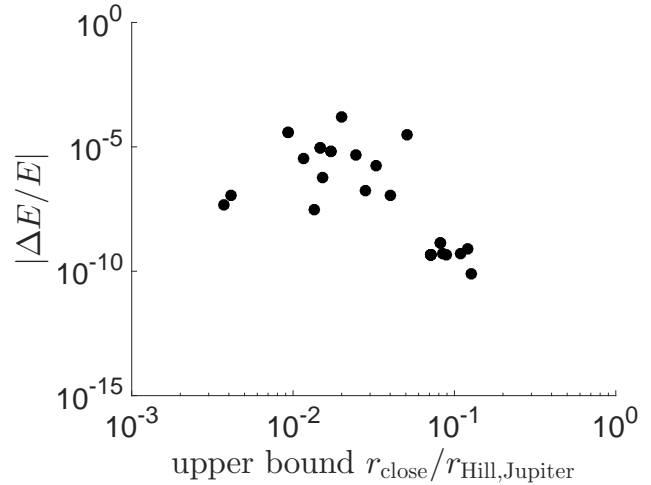}
  \caption{Energy error as a function of the closest approach for the solution of the massive outer gas giants problem.  A solution is obtained for the first 3000 years using HB15 using step size $h = 0.005$ years.  The close encounter separation is an upper bound because it is calculated at the end of time steps.}
  \label{fig:phier}
  \centering
\end{figure}

In Fig. \ref{fig:phier} we plot the error at the end of 3000 years as a function of the smallest separation between two particles during the run.  Distances between particles are calculated at the end of each time step, so the smallest separations in Fig. \ref{fig:phier} are actually an upper bound of the minimum separation which could occur during time steps.  The distance is in units of the massive Jupiter's Hill radius at time 0, 1.15 AU.  Jupiter's Hill radius at time 0 is smaller than the other planets' Hill radii.  HB15 integrated all initial conditions with moderately small error for inferring statistics of the $N$-body problem \citep{PB14}.  For integrations of many Lyapunov times, high accuracy solutions probably are inefficient.  The error does not depend clearly on the smallest separation; a poor linear fit to the points give a slope $\sim -3.3$.  The smallest close encounter upper bound in the plot is $r_{\mathrm{close}} = 0.0043$ AU, which is smaller than the minimum separation found by \cite{C99}.  HB15 can handle a variety of close encounter situations.  It has been reported that the method of \cite{DLL98} fails for close enough encounters \citep{K00}.  
\section{Limitations of \texttt{MERCURY} and Switching Integrators}
\label{sec:switch}
\subsection{Switching Integrators for the $N$-body problem}

Authors that have solved the $N$-body problem with switching integrators include \cite{C99}, \cite{DLL98}, \cite{OFM11}, \cite{IPM15}, and \cite{K00}.  The common strategy is to switch the method to a more suitable choice when a pair of particles comes in close proximity.

\citeauthor{K00} studied the effect of the smoothness of the switch in a short timescale problem: 32 orbits of a two-body problem.  Their integrator switches between two different symplectic methods.  The switch can occur gradually or suddenly: the transition is quantified by a smoothing function.  The integrator takes constant time steps.  For the $C^0$ function, a theta step function $\theta(r)$, with $r$ the particle pair separation, the switch occurs at different phases along the orbit, once the integrator realizes it is in a new method region.    The $C^0$ function caused clear secular growth in energy error.  As the smoothness was increased from $C^0$ to $C^8$, the secular growth was gradually mitigated, but presumably not eliminated.  They also found deterioration in time reversibility due to the switches. 

Symplectic maps have an associated function $\tilde{H}$, which is an integral when it converges \citep{DF76}.  Thus, it is analogous to a Hamiltonian. $\tilde{H}$ is time independent if the time step is not varied as a function of phase space.  To understand the effects of switching integrators, we discuss $\tilde{H}$ further using the Standard Map \citep{LL83}, a problem with two canonical variables, $x$ and $p$, as an example.  With $K$ positive, the Standard Map is
\begin{equation}
\begin{aligned}
p^{\prime} &= p + \frac{K}{2 \pi} \sin(x) \\
x^{\prime} &= x + p^{\prime}. \\
\end{aligned}
\label{eq:stmap}
\end{equation}
$x$ and $p$ are mapped to $x^{\prime}$ and $p^\prime$.  The question is what is the continuous-time Hamiltonian corresponding to eq. \ref{eq:stmap}.  There are two possible answers.  The Standard Map is closely related to the one degree of freedom simple pendulum problem with Hamiltonian
\begin{equation}
H(x,p) = \frac{p^2}{2} + \frac{K}{2 \pi} \cos (x),
\label{eq:pend}
\end{equation}
which is integrable and non-chaotic.
For the first answer, add a time dependence to equation \ref{eq:pend}, $H_1 = K/(2 \pi) \sum_{n \ne 0} \cos (x - n \Omega t)$.  $\Omega$ is a positive constant and $n$ an integer.  Using an identity for Dirac delta functions, we get as the sum $H + H_1$ a new time dependent Hamiltonian
\begin{equation}
H_{\delta}(x,p,t) = \frac{p^2}{2} + \frac{K}{2 \pi} \cos(x) \sum_n \delta(\Omega t - 2 \pi n).
\label{eq:hdel}
\end{equation}
\cite{C79} first wrote down eq. \ref{eq:hdel}.  Let $\Omega = 2 \pi$.  Then solving Hamilton's equations for $H_{\delta}$ starting from $t = 0$ and integrating to $t = 1$ leads to map \ref{eq:stmap}.  Thus the first Hamiltonian corresponding to map \ref{eq:stmap} is the time dependent $H_{\delta}$.  

The second answer is a result of an approximation to solving Hamiltonian \ref{eq:pend}.  Map \ref{eq:stmap} also results from applying a time independent Hamiltonian $\tilde{H}$ for time $h$  \citep{hair06}:
\begin{equation}
\begin{aligned}
\tilde{H}(x,p) &= H(x,p) - \frac{h K}{2 \pi} p \sin(x) + \frac{h^2 K}{24 \pi} \left( \frac{K}{2 \pi} \sin^2(x) - p^2 \cos(x) \right)\\
&  + \mathcal O(h^3). 
\end{aligned}
\label{eq:htsm}
\end{equation}
$\tilde{H}$ arises from applying symplectic Euler as a splitting method to Hamiltonian \ref{eq:pend}.  It can be derived following \cite{HB15}.  When we let $h = 1$ in eq. \ref{eq:htsm}, we recover eq. \ref{eq:stmap}.  For $K$ small, $p$ is bounded, and we find the phase space is constrained by contours of $\tilde{H}$ written to third order in $h$, a result also found by \cite{Y93}.  This is a result consistent with preservation of $\tilde{H}$.  We do not expect $\tilde{H}$ to converge for chaotic orbits because the orbits do not have an isolating integral.  The solutions of applying $\tilde{H}$ and $H_{\delta}$ coincide, if $\Omega = 2 \pi/h$, at time $t = h$.  We focus on the effect of a switching integrator on $\tilde{H}$, not $H_{\delta}$ because it is independent of a continuous time variable.

\subsection{A Switch in the Simple Harmonic Oscillator}
\label{sec:shoswitch}

We focus on an integrable, one-degree of freedom problem, the simple harmonic oscillator (SHO).  We use it because of its simple regular behavior.  The Hamiltonian is
\begin{equation}
H(x,p) = \frac{p^2 + x^2}{2}.
\label{eq:SHO}
\end{equation}
The period $P$ of the motion described by $H$ is $2 \pi$.  Let $y(t)$ be the phase space vector $(x,p)$ at time $t$.  An exact solution for $y(t)$ is found from
\begin{equation}
y(t) = \exp\left( t D_H \right) y(0).
\end{equation}
The operator $D_H$ is defined by $D_H y(t) = \left\{y(t),H \right\}$, where $\{\}$ are Poisson brackets.  Because eq. \ref{eq:SHO} is integrable, there is no need to construct a switching method to solve it, but we do so because it will reveal useful properties of switching integrators.  Rewrite eq. \ref{eq:SHO} as $H = H_1 + H_2$ with $H_1 = \left(p^2 + x^2 K(x)\right)/2$ and $H_2 = x^2 \left(1 - K(x)\right)/2$.  Then an approximation to $y(t)$ is
\begin{equation}
\tilde{y}(t) = \exp(t D_{H_1}) \exp(t D_{H_2}) y(0).
\label{eq:switch}
\end{equation}
For $K = 0$, eq. \ref{eq:switch} gives the symplectic Euler solution.  If we choose $t=h<1$ and iterate eq. \ref{eq:switch} with $K = 0$, the solution trajectory satisfies the equation of an ellipse with eccentricity $e < 1$, $A p^2 + B p x + C x^2 - 1 = 0$, with $A$, $B$, and $C$ positive numbers.  $A = A(h,\tilde{H}_{\mathrm{SO}})$, $B = B(h)$, and $C = C(h,\tilde{H}_{\mathrm{SO}})$, where $\tilde{H}_{\mathrm{SO}}$ is the function analogous to eq. \ref{eq:htsm} for the Hamiltonian of the SHO.  As $h \rightarrow 0$, $e \rightarrow 0$.  For $K = 1$, iterating eq. \ref{eq:switch} yields the circular $e = 0$ solution trajectory of the SHO Hamiltonian.  If $h \rightarrow 0$ for the $K = 0$ trajectory, the two $K = 1$ and $K = 0$ solution trajectories coincide.     

\begin{figure}
 \includegraphics[width=0.5\textwidth]{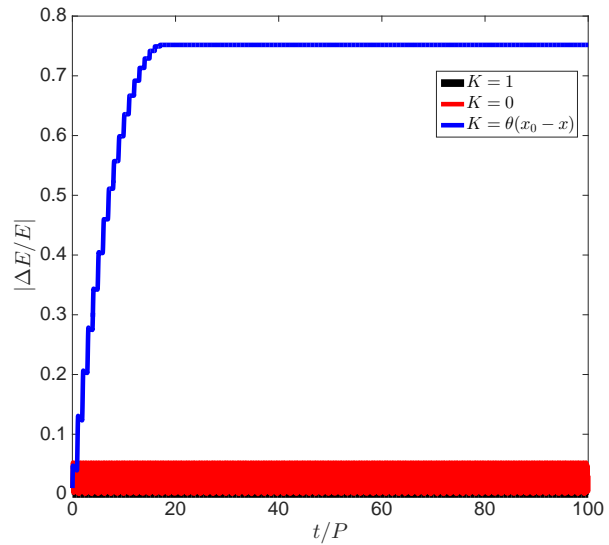}
  \caption{Energy error as a function of time for the solution of the SHO problem.  We use a switching method with $h = 0.1$ to obtain the solution.  We plot the $K = 0$, $K = 1$, and $K = K(x)$ solutions.  The $K = 1$ solution hugs the $x$-axis.  The $K = K(x)$ method shows a secular increase in error due to time dependence introduced into $\tilde{H}_{\mathrm{switch}}$.}
  \label{fig:swtch}
  \centering
\end{figure}

As a simple example, let $K$ be a $C^0$ function, $K(x) = \theta (x_0 - x)$: $K(x)$ is either 0 or 1.  We choose $p(0) = 0$, $x(0) = 1$, and $x_0 = 0.5$.  The $K = 1$ solution in the $x-p$ plane describes a circle of radius 1.  We run the problem for $100 P$.  We use a step size $h = 0.1$ so there are $\approx 63$ steps per period.  We plot the error in the energy as a function of time in Fig. \ref{fig:swtch}.  We also plot the $K = 1$ solution (it is hugging the $x$-axis) and the $K = 0$ solution.    
We see that the $K = 0$ solution has no secular drift in error, except for roundoff error.  Technically, the growth of energy error for the $K = 1$ solution, not visible in Fig. \ref{fig:swtch}, is linear in time: this $t^1$ dependence of the error is larger than the $t^{1/2}$ growth described by Brouwer's law.  The $K = 0$ solution has a a secular error drift that is exponentially suppressed and negligible in this case \citep{hair06}.  

For $K = \theta (x_0 - x)$ there is a severe growth in error.  The growth appears to stop shortly before $t/P = 20$.  The explanation for this behavior is as follows.  As $t$ increases, $H$ and $\tilde{H}_{\mathrm{SO}}$ increase by a factor $2$ or more as the maximum $|p|$ and $|x|$ increase.  The difference between $H$ and $\tilde{H}_{\mathrm{SO}}$ becomes smaller than the difference of either of them with the initial energy and the growth of the blue curve diminishes.  If we reduce $h$, the growth asymptotes at large $t$.  For larger $h$, the growth asymptotes at smaller $t$.  We can measure the dependence on $h$ of the energy error after $100 P$.  For small enough $h$, $h \le 2 \times 10^{-3} P$, $\Delta E/E$ depends linearly on $h$, as expected for a first order method.  We tried higher order switches and different $x_0$ values and our results are consistent with the results of \cite{K00}.

We can write $\tilde{H}_{\mathrm{switch}}$ for the solution of eq. \ref{eq:switch} by casting eq. \ref{eq:switch} into the form
\begin{equation}
\tilde{y}(h) = \exp(h D_{\tilde{H}_{\mathrm{switch}}}) y(0).
\end{equation}
This equation defines $\tilde{H}_{\mathrm{switch}}$.  $\tilde{H}_{\mathrm{switch}}$ is then (for more details in the derivation, see \cite{HB15})
\begin{equation}
\begin{aligned}
\tilde{H}_{\mathrm{switch}}(x,p) &= \frac{p^2 + x^2}{2} + h \left(-x p + \frac{\theta(x-x_0) p}{2} + \frac{\delta(x-x_0) x^2 p}{2} \right) \\
& + \mathcal O (h^2).
\end{aligned}
\label{eq:disc}
\end{equation}
The discontinuities in phase space coordinate derivatives of $\tilde{H}_{\mathrm{switch}}$ are responsible for the break in symplecticity of the harmonic oscillator switch.  As expected $\tilde{H}_{\mathrm{switch}}$ is time independent, which seems to contradict the secular growth in error observed in Fig. \ref{fig:swtch}.  The solution to this apparent contradiction is that we introduced time dependence through the behavior at coordinates $x$ near $x_0$.  Because in practice we switch at $x$ different from $x_0$, within $\approx h/(2 \pi)$ of $x_0$, we are changing the integrator as a function of time, and we are making $\tilde{H}_{\mathrm{switch}}$ time dependent.  A similar effect with the \texttt{MERCURY} solution was undetectable in our tests.  This may be due to the high smoothness of the smoothing function \citep{K00}.  \texttt{MERCURY} introduces additional time dependence to its $\tilde{H}$ because its $r_{\mathrm{crit}}$ is a function of time, and the smoothing function is a function of $r_{\mathrm{crit}}$.  Any symplectic switching method that does not carefully solve boundary behavior in the switch will suffer from time dependence in its $\tilde{H}$, possibly causing detectable secular energy error growth in time.  
\subsection{Symplecticity of \texttt{MERCURY}}
\label{sec:symp}
Switching integrators that switch between two symplectic methods are not symplectic.  The integrator is not symplectic at the phase space points where the smoothing function joins the two methods.  $\tilde{H}$ has discontinuities at these phase space points because we cannot practically construct a $C^{\infty}$ function that is infinitely differentiable.  An example of such a discontinuous $\tilde{H}$ is shown by eq. \ref{eq:disc}.

Nonetheless, the problem points are rare and we expect nearby initial conditions to preserve the Poincar\'{e} invariants when mapped by a switching method like \texttt{MERCURY}.  Because we are unaware of rigorous results of the symplecticity of a switching method, we test \texttt{MERCURY}'s symplecticity by calculating the $6N \times 6N$ equations in the symplectic condition \citep{SW01}, 
\begin{equation}
\Omega = J^{\dag} \Omega J.
\label{eq:symp}
\end{equation}
$\Omega$ is a constant matrix if we work in a canonical coordinate system.  Its form depends on the ordered basis.  For ordered basis (x,p), $\Omega$ is
\begin{equation}
{\Omega} = 
	\begin{bmatrix}
	{0} & {{I}} \\
	-{{I}} & {0} \\
	\end{bmatrix}.
\end{equation}
$I$ is the $3N \times 3N$ identity matrix.  $J$ is the Jacobian of the map, and $J^\dag$ its transpose.  Eq. \ref{eq:symp} is satisfied if and only if the method is symplectic.

We take one time step of length $h = 0.001$ years in the binary planet problem of Section \ref{sec:hier}.  $h = 0.001$ years is $3 \%$ of the binary planet orbital period.  We calculate $dR$, which is the sum of the absolute value of all elements in the remainder matrix $R = J^{\dag} \Omega J - \Omega $: it's $L_1$ norm, with no normalization.  For a symplectic method $dR$ should be 0 neglecting errors in estimating derivatives and finite precision arithmetic.  The summation in this norm is over elements of different units.  A norm that sums elements of the same units consists of summing the absolute values of elements along the diagonal from the lower left to upper right of $R$.  This alternate norm did not effect our values of $dR$ significantly.  This alternate norm can be a useful way to quantify symplecticity which we plan to investigate in future work.    

We compute Jacobian elements in eq. \ref{eq:symp} by calculating $8$th order finite differences as a proxy for partial derivatives.  For each Jacobian column we calculate, we run a method 8 times.  There are 12 columns corresponding to 12 phase space variables.  So to calculate $J$, we run a method $96$ times.  We test our calculation of $dR$ on various well known methods.  In Table \ref{tab:symplec} we record the values of $dR$ and the typical energy error for the close initial conditions used in calculating $dR$. 
\begin{center}
\begin{table}
\caption{Measure of symplecticity $dR$ for the solutions of the binary planet hierarchical triple problem.  Various integrators are used with displaced initial conditions and $h = 0.001$ years to obtain solutions.  Also in the table is the typical error in energy for the initial conditions used in calculating $dR$.  The three formally symplectic methods, WHJ, HB15, and DKD leapfrog, have the lowest $dR$.}
\centering
\begin{tabular}{| c || c| c|}
	\hline
	 Method & $dR$ &$\left|\Delta E/E \right|$  \\ [3ex] \hline
	WHJ &  $5.2\times 10^{-10}$  &  $1\times 10^{-7}$\\ 
	BS &  $8.6\times 10^{-10}$ &  $3\times 10^{-16}$\\ 
	Radau & $5.6\times 10^{-10}$  &  $1\times 10^{-16}$\\ 
	HB15 & $8.8\times 10^{-11}$  &  $3\times 10^{-10}$\\ 
	DKD Leapfrog & $3.9\times 10^{-11}$&  $4\times 10^{-7}$  \\ 
	\texttt{SAKURA} & $1.1\times 10^{-9}$ &  $1\times 10^{-4}$\\ 
	RK4 & $2.2\times 10^{-6}$ & $3\times 10^{-10}$ \\ \hline
	\end{tabular}
\label{tab:symplec}
\end{table}
\end{center}
We test three integrators in the \texttt{MERCURY} software package: WHJ with symplectic correctors, BS, and Radau.  WHJ is the only strictly symplectic method, but BS and Radau behave almost as well as a symplectic integrator.  Radau behaves like a symplectic integrator due to the effects of finite precision and roundoff error.  Symplecticity as a concept is most useful for real valued functions when roundoff error is ignored.  The small $dR$ is reasonable for BS as BS is highly accurate with an accuracy tuned by Tol.  We show $dR$ for four other integrators not part of the \texttt{MERCURY} software: symplectic methods HB15 and Drift-Kick-Drift (DKD) leapfrog (see for example \citep{HB15}), and non-symplectic methods \texttt{SAKURA} \citep{GBP14} and 4th order Runge-Kutta (RK4).  The non-symplectic methods \texttt{SAKURA} and RK4 have the largest $dR$.  

We note that the magnitude of energy errors are not indicative of symplecticity.  Two of the three methods with worst energy error, leapfrog and WHJ, are also two of the three methods that best conserve Poincar\'{e} invariants.  By contrast, RK4 has a small energy error, similar to HB15's error, yet RK4 has the largest phase space volume element conservation error from all methods.  Errors in first integrals do not indicate how well phase space volume elements are conserved.

\begin{figure}
 \includegraphics[width=.5\textwidth]{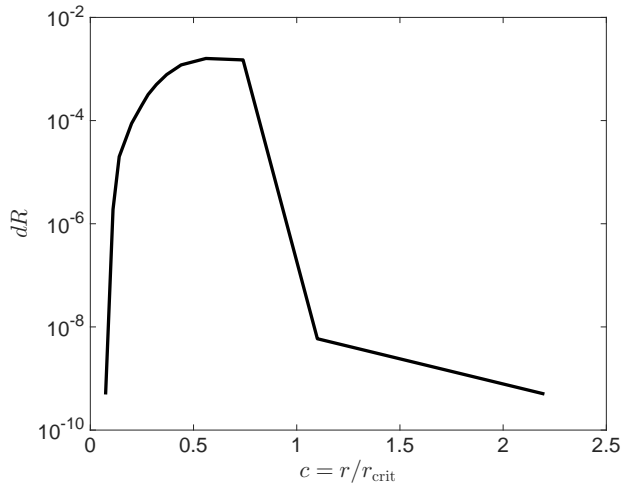}
  \caption{A measure of symplecticity for a phase space tube of \texttt{MERCURY} solutions of the binary planet hierarchical triple problem.  $h = 0.001$ is used.  The symplecticity deteriorates when $c$ is in the range 0.1 to 1.}
  \label{fig:mercsymp}
  \centering
\end{figure}

Next we test the \texttt{MERCURY} switching integrator.  We vary $r_{\mathrm{crit}}$, which measures where in phase space the switch takes place.  We calculate $dR$ for each value of $c = r/r_{\mathrm{crit}}$, where $r$ is the initial separation between the binary planets.  Subject to other phase space coordinate requirements \citep{C99}, for $c < 0.1$, \texttt{MERCURY} reduces to BS, which we showed yields small $dR$.  WHD.  For $c > 1$, \texttt{MERCURY} reduces to WHD, which is a known symplectic method.    The result is shown in Fig. \ref{fig:mercsymp}.  
We find the result that phase space volume element conservation is strongly violated for a range of values of $c$ when testing the binary planet problem of \cite{DLL98}.  Switching methods are predicted to be non-symplectic, but the large violation we found may be due to other reasons, such as the code's close encounter predictor between time steps.  
\subsection{Time Reversibility of \texttt{MERCURY}}
\label{sec:trevers}
Time reversibility, or time symmetry, is a highly desirable property of an integrator.  \cite{hair06} show it can lead to benefits akin to symplecticity.  To test reversibility of a method, we can run a method for time $h$ and then for time $-h$.  We should recover the initial conditions, neglecting roundoff error, regardless of the magnitude of $h$.  

WHJ is reversible, as we can see analytically and by numerical tests.  We see analytically that WHD is also reversible.  Radau and BS are approximately time symmetric in our tests.  We test the reversibility of \texttt{MERCURY} by running forwards and backwards the hierarchical binary planet problem of Section \ref{sec:hier} for several positive values of $h$ and calculating the energy error.  We try different values of $r_{\mathrm{crit}}$.  For reference, we also test HB15, which is a reversible method for a perfect Kepler solver, ignoring roundoff error.  We test HB15 with the \cite{WH15} and $\texttt{drift\_one}$ Kepler solvers.

\begin{figure}
 \includegraphics[width=.5\textwidth]{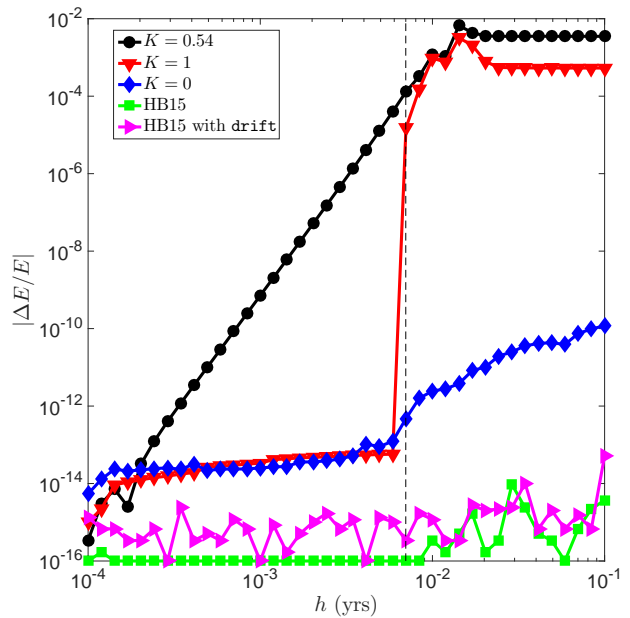}
  \caption{Time reversibility of \texttt{MERCURY} and HB15.  A step forwards and then a step backwards of various sizes is taken when solving the hierarchical three body problem.  Reversibility breaks strongly for the \texttt{MERCURY} $K = 1$ and $K = 0.54$ points.}
  \label{fig:trevers2}
  \centering
\end{figure}

The results are shown in Fig. \ref{fig:trevers2}.
We plot reversibility curves for different values of the smoothing function $K$, which ranges from 0 (Bulirsch Stoer) to 1 (WHD).  To use a middle $K = 0.54$, we choose $r_{\mathrm{crit}} = 0.26 r_{\mathrm{Hill}}$.   $r_{\mathrm{Hill}} = a (m/(3M))^{1/3}$ with $a = 1$ AU, $m = 10^{-3} M_{\odot}$ and $M = 1 M_{\odot}$.  We again input $c$ (Section \ref{sec:symp}) into \texttt{MERCURY} and calculate $K$ by using eq. 10 in \cite{C99}.  All curves show breaks in reversibility; we must pick apart breaks due to roundoff or inherent to the methods.  Errors in HB15 are seen to be due in part to the Kepler solver.  The other part of the errors are due to roundoff error.  BS and WHD only weakly break time reversibility until $h/P = 0.2121 = h_0$, where $P$ is the binary planet period.  The forward time step brings the binary closer to periapses as $h$ increases in Fig. \ref{fig:trevers2}.  For $h > h_0$ \texttt{MERCURY} establishes a velocity dependent $r_{\mathrm{crit}}$ in the $K = 1$ points which causes a large reversibility break.  The error of the $K = 0$ points grows more rapidly also for $h > h_0$.  The $K = 0.54$ curve has a close encounter at every time step.  It grows as $h^6$ initially.  The BS tolerance is set to $10^{-12}$.  Even though the two integrators of \texttt{MERCURY} used individually are approximately reversible, used in combination, in \texttt{MERCURY}, they are not.  

\subsection{Dynamically wrong results from \texttt{MERCURY}}
Sections \ref{sec:shoswitch}, \ref{sec:symp}, and \ref{sec:trevers} have highlighted concerns with \texttt{MERCURY} and other switching methods.  In this Section, we show an example where \texttt{MERCURY} gives dynamically incorrect results.  The wrong results appear to be related to a bug because they persist across various integrators in the \texttt{MERCURY} software package.

We again consider the binary planet problem of Section \ref{sec:hier}.  We increase the binary eccentricity to $e = 0.98$.  The planet masses are each reduced by $10\%$ to $8.9 \times 10^{-4} M_{\odot}$.  We set $r_{\mathrm{crit}} = 3 r_{\mathrm{Hill}}$ in an attempt to induce two switches per period $P$ of the binary planet orbit near its periapses: during its orbit, its $c$ (Section \ref{sec:symp}) will range from $1.2 \times 10^{-4}$ at apoapse to $0.12$ at periapses.  This test problem is artificial and was chosen to induce frequent switches. 

We run \texttt{MERCURY} and HB15 for 1,000 years with different step sizes and tolerances (Tol) and measure the energy error and the unbinding time $t_U$ of the binary planet.  We show results for \texttt{MERCURY} in Table \ref{tab:unbindmerc} and results for HB15 in Table \ref{tab:unbindhb15}.  The difference is that $t_U < 25$ years in all \texttt{MERCURY} runs, while $t_U > 1000$ years in all HB15 runs.  We investigated a solution from the final row of Table \ref{tab:unbindmerc} and found $t_U$ corresponds approximately to a periapses passage.  Despite the small errors in energy and angular momentum (now shown) in \texttt{MERCURY}, we show evidence that the correct $t_U$ is given by the phase space volume conserving HB15 method.  

The periodic behavior of the phase space coordinates suggests the HB15 solutions are approximately quasi-periodic for at least 1000 years, and the \texttt{MERCURY} solutions are approximately quasi-periodic until their time $t_U$.  To study this, we focus attention on the $h/P = 9.2 \times 10^{-4}$ and Tol = $1 \times 10^{-12}$, for \texttt{MERCURY}, solutions.  For one method at a time, we calculate the $L_1$ norm as a function of time between solutions in a phase space tube that have nearby initial conditions.  $L_1$ as a function of time quantifies the divergence of the solutions.  At least until 200 years, HB15's solutions have no clear indication of exponential divergence.  Until its $t_U$, \texttt{MERCURY} has no clear indication of exponential divergence either.  We find no clear evidence of exponential divergence, characteristic of chaotic dynamics, for the solution trajectories of either method.  We visually inspect trajectories by choosing one of the 12 phase space coordinates, and plotting it as a function of time for both HB15 and \texttt{MERCURY}, on the same graph.  The $y$-axis scale is set by the maximum and minimum phase space coordinate values between time 0 and $ 3.9$ years, \texttt{MERCURY}'s $t_U$.  We find the HB15 and \texttt{MERCURY} trajectories are indistinguishable until time $3.9$ years.  We repeated the experiment for several phase space coordinates and the trajectories in time remained indistinguishable.  The results for HB15 and \texttt{MERCURY} are consistent with quasiperiodic, non-chaotic, motion.       

\begin{center}
\begin{table}
\caption{Time of binary unbinding $t_U$ of \texttt{MERCURY} for several time steps and tolerances.  We did not find a set of parameters for which $t_U$ was larger than 25 years.  We also show the small energy error after 1000 years.}
\centering
\begin{tabular}{| c || c| c| c|}
	\texttt{MERCURY} & & & \\ \hline
	$h/P$ & Tol & $t_U$/years & $\left|\Delta E/E\right|$ \\ [3ex] \hline
	$9.2\times 10^{-2}$ &  $1\times 10^{-12}$  &  11 &  $1.3 \times 10^{-9}$ \\ 
	$9.2\times 10^{-3}$ & $1\times 10^{-14}$  & 25 &  $1.9\times 10^{-9}$ \\ 
	$9.2\times 10^{-4}$ & $1\times 10^{-11}$ and $1\times 10^{-12}$& 3.9 &  $1.2\times 10^{-9}$ \\ \hline
	\end{tabular}
\label{tab:unbindmerc}
\end{table}
\end{center}
\begin{center}
\begin{table}
\caption{Time of binary unbinding $t_U$ of HB15 for different time steps.  We did not find time steps for which $t_U$ was less than 1000 years.}
\centering
\begin{tabular}{|| c| c| c|}
	HB15 & & \\ \hline
	$h/P$ &$t_U$/years & $\left|\Delta E/E\right|$ \\ [3ex] \hline
	$9.2\times 10^{-3}$ &  $> 1000$ &  $3.5 \times 10^{-4}$ \\ 
	$1.8\times 10^{-3}$ & $> 1000$ &  $3.7\times 10^{-7}$ \\ 
	$9.2\times 10^{-4}$ & $> 1000$ &  $4.5\times 10^{-10}$ \\ \hline
	\end{tabular}
\label{tab:unbindhb15}
\end{table}
\end{center}
To confirm HB15's results for $t_U$, the same test was run with a different widely used integrator, not part of the \texttt{MERCURY} software, the fourth order Hermite integrator with shared adaptive time steps, described in \cite{K98}.  The method is claimed to be approximately time symmetric for near constant time steps.  In one comparison below, it is significantly slower than HB15.  Different $\eta$ \citep[Chapter 1]{aar08} and iteration numbers are set and the average $h$, $<h>$, is calculated.  Some results are shown in Table \ref{tab:unbindherm}. 

\begin{center}
\begin{table}
\caption{Time of binary unbinding $t_U$ of a time symmetric Hermite method for several values of $\eta$ and number of iterations.  No set of parameters were found for which $t_U$ was less than 1000 years.}
\centering
\begin{tabular}{|| c | c || c| c| c|}
	Hermite & & & & \\ \hline
	$<h>/P$ & $\eta$ & Max. Iter. & $t_U$/years & $\left|\Delta E/E\right|$ \\ [3ex] \hline
	$3.2\times 10^{-3}$ &  0.1 & 5 &  $>1000$ &  $4.8 \times 10^{-4}$ \\ 
	$1.6\times 10^{-3}$ & 0.05  & 3 & $>1000$ &  $1.6\times 10^{-5}$ \\ 
	$3.1\times 10^{-4}$ & 0.01 & 5 & $>1000$ &  $8.0\times 10^{-9}$ \\ \hline
	\end{tabular}
\label{tab:unbindherm}
\end{table}
\end{center}
$t_U$ is again always $>1000$ years.  For $\eta = 0.05$, a precession in the orientation of the Runge-Lenz vector of the binary was measured with period of $\approx 329$ years; a similar effect is seen when two body eccentric orbits are studied with leapfrog.  Leapfrog shows precession because there is no nearby first integral to the argument of pericenter in the exact two-body leapfrog solution.  The two-body leapfrog solution approximately conserves the actions, responsible for the solution topology, but has linear drift in the angles error.  The third row of Table \ref{tab:unbindherm} took 6-10 times more computing effort than the third row of Table \ref{tab:unbindhb15}, despite the HB15 error being smaller.  We confirmed the results of HB15 and Hermite using an explicit 4th order symplectic method from \cite{Y90} and an Adams-Bashforth-Moulton predictor-corrector linear multistep method \cite{BT08}.

Thus, we have found an artificial $N$-body problem for which \texttt{MERCURY} predicts a binary will unbind, unphysically, despite \texttt{MERCURY}'s small errors in energy and angular momentum.  We included print statements, within the \texttt{MERCURY} software to verify the ejections taking place with positions of several hundred AU.  The Radau and BS methods in the \texttt{MERCURY} package also had $t_U < 50$ years, and behave incorrectly.     
\section{Conclusions and Recommendations}
\label{sec:conc}
For years, switching methods like \texttt{MERCURY} and \texttt{SyMBA} have been popular methods for solving problems in planetary system dynamics and formation.  \texttt{MERCURY} is cited frequently, is publicly available, and easy to use.  The first aim of this paper is to describe limitations of switching methods in general through a toy problem.  We then show that methods in the \texttt{MERCURY} software package can give dynamically incorrect results even when first integral errors are small.  We suggest caution must be exercised when using \texttt{MERCURY} methods and solutions should be checked using other methods.

We apply the exactly - ignoring roundoff and Kepler solver error- symplectic and reversible method of \cite{HB15} that uses intuitive inertial Cartesian coordinates to problems in planetary system dynamics and offer it as an alternative method for planetary dynamics simulations.  For an adequate time step, it was able to integrate accurately every close encounter in its solutions.  When compared against other methods in \texttt{MERCURY} software, it is found to be the most efficient method, or tied as the most efficient method in the moderate error range for a number problems.  Further efficiency tests were done in \cite{HB15}.  

HB15 is useful for long term studies, when high accuracy solutions by high order methods may be unnecessarily expensive.  The errors of HB15 could be reduced by orders magnitude by using a symplectic corrector and its inverse at virtually no additional computational cost.  We have not developed symplectic correctors here; because they will involve three, and not two, generally non-commuting operators, coefficient formulae in \cite{W06} do not apply.  

HB15 is easy to implement and its steps are described partly in Algorithm \ref{alg:phih}.  We recommend its use with the Kepler solver of \cite{WH15}.  It is designed for solving general planetary system dynamics problems including those with multiple stars, binary planets, or planets with satellites.  

\section{Acknowledgements}
This work was begun after conversations with Piet Hut, Gerry Sussman, and Scott Tremaine.  I appreciate feedback from Ed Bertschinger, Jack Wisdom, Scott Tremaine, and Katherine Deck.  The results of the Hermite code were obtained by Kento Masada, and results of the Adams-Bashforth-Moulton method were obtained by Uchupol Ruangsri.  I thank the anonymous referee for detailed helpful comments.  I acknowledge support by the National Science Foundation Graduate Research Fellowship under Grant No. 1122374.

\bibliographystyle{mn2e}
\bibliography{doc}

\begin{thebibliography}{40}
\expandafter\ifx\csname natexlab\endcsname\relax\def\natexlab#1{#1}\fi

\bibitem[{{Aarseth}, {Tout} \& {Mardling}(2008){Aarseth}, {Tout}, \&
  {Mardling}}]{aar08}
{Aarseth} S.~J., {Tout} C.~A., {Mardling} R.~A., eds., 2008, Lecture Notes in
  Physics, Berlin Springer Verlag, Vol. 760, {The Cambridge N-Body Lectures}.
  Springer Verlag, Berlin

\bibitem[{{Batygin}, {Brown} \& {Betts}(2012){Batygin}, {Brown}, \&
  {Betts}}]{B12}
{Batygin} K., {Brown} M.~E., {Betts} H., 2012, ApJL, 744, L3

\bibitem[{{Binney} \& {Tremaine}(2008)}]{BT08}
{Binney} J., {Tremaine} S., 2008, {Galactic Dynamics: Second Edition}.
  Princeton University Press

\bibitem[{{Brouwer}(1937)}]{B37}
{Brouwer} D., 1937, AJ, 46, 149

\bibitem[{{Chambers}(1999)}]{C99}
{Chambers} J.~E., 1999, MNRAS, 304, 793

\bibitem[{{Chambers} {et~al}\mbox{.}(2002){Chambers}, {Quintana}, {Duncan}, \&
  {Lissauer}}]{C02}
{Chambers} J.~E., {Quintana} E.~V., {Duncan} M.~J., {Lissauer} J.~J., 2002, AJ,
  123, 2884

\bibitem[{{Channell} \& {Scovel}(1990)}]{chan90}
{Channell} P.~J., {Scovel} C., 1990, Nonlinearity, 3, 231

\bibitem[{{Chirikov}(1979)}]{C79}
{Chirikov} B.~V., 1979, Physics Reports, 52, 263

\bibitem[{{Danby}(1988)}]{dan88}
{Danby} J.~M.~A., 1988, {Fundamentals of celestial mechanics}, 2nd edn.
  Willmann-Bell, Richmond, Va., U.S.A.

\bibitem[{{Dragt} \& {Finn}(1976)}]{DF76}
{Dragt} A.~J., {Finn} J.~M., 1976, Journal of Mathematical Physics, 17, 2215

\bibitem[{{Duncan}, {Levison} \& {Lee}(1998){Duncan}, {Levison}, \&
  {Lee}}]{DLL98}
{Duncan} M.~J., {Levison} H.~F., {Lee} M.~H., 1998, AJ, 116, 2067

\bibitem[{{Everhart}(1985)}]{E85}
{Everhart} E., 1985, in Dynamics of Comets: Their Origin and Evolution,
  Proceedings of IAU Colloq. 83, held in Rome, Italy, June 11-15, 1984. Edited
  by Andrea Carusi and Giovanni B. Valsecchi. Dordrecht: Reidel, Astrophysics
  and Space Science Library. Volume 115, 1985, p.185, {Carusi} A., {Valsecchi}
  G.~B., eds., p. 185

\bibitem[{{Gon{\c c}alves Ferrari}, {Boekholt} \& {Portegies
  Zwart}(2014){Gon{\c c}alves Ferrari}, {Boekholt}, \& {Portegies
  Zwart}}]{GBP14}
{Gon{\c c}alves Ferrari} G., {Boekholt} T., {Portegies Zwart} S.~F., 2014,
  MNRAS, 440, 719

\bibitem[{{Gustafson}(2015)}]{G15}
{Gustafson} J.~L., 2015, {The End of Error: Unum Computing}. Chapman and
  Hall/CRC

\bibitem[{{Hairer}, {Lubich} \& {Wanner}(2006){Hairer}, {Lubich}, \&
  {Wanner}}]{hair06}
{Hairer} E., {Lubich} C., {Wanner} G., 2006, {Geometrical Numerical
  Integration}, 2nd edn. Springer Verlag, Berlin

\bibitem[{{Hairer}, {McLachlan} \& {Razakarivony}(2008){Hairer}, {McLachlan},
  \& {Razakarivony}}]{HMR07}
{Hairer} E., {McLachlan} R., {Razakarivony} A., 2008, BIT, 48, 231

\bibitem[{{Heggie} \& {Hut}(2003)}]{HH03}
{Heggie} D., {Hut} P., 2003, {The Gravitational Million-Body Problem: A
  Multidisciplinary Approach to Star Cluster Dynamics}. Cambridge University
  Press

\bibitem[{{Henrici}(1962)}]{H62}
{Henrici} P., 1962, {Discrete variable methods in ordinary differential
  equations}

\bibitem[{{Hernandez} \& {Bertschinger}(2015)}]{HB15}
{Hernandez} D.~M., {Bertschinger} E., 2015, MNRAS, 452, 1934

\bibitem[{{Iwasawa}, {Portegies Zwart} \& {Makino}(2015){Iwasawa}, {Portegies
  Zwart}, \& {Makino}}]{IPM15}
{Iwasawa} M., {Portegies Zwart} S., {Makino} J., 2015, Computational
  Astrophysics and Cosmology, 2, 6

\bibitem[{{Kokubo}, {Yoshinaga} \& {Makino}(1998){Kokubo}, {Yoshinaga}, \&
  {Makino}}]{K98}
{Kokubo} E., {Yoshinaga} K., {Makino} J., 1998, MNRAS, 297, 1067

\bibitem[{{Kvaerno} \& {Leimkuhler}(2000)}]{K00}
{Kvaerno} A., {Leimkuhler} B., 2000, SIAM J. Sci. Comp., 22, 1016

\bibitem[{{Laskar} {et~al}\mbox{.}(2004){Laskar}, {Robutel}, {Joutel},
  {Gastineau}, {Correia}, \& {Levrard}}]{L04}
{Laskar} J., {Robutel} P., {Joutel} F., {Gastineau} M., {Correia} A.~C.~M.,
  {Levrard} B., 2004, AAp, 428, 261

\bibitem[{{Lichtenberg} \& {Lieberman}(1983)}]{LL83}
{Lichtenberg} A.~J., {Lieberman} M.~A., 1983, {Regular and stochastic motion}.
  Springer Verlag

\bibitem[{{Oshino}, {Funato} \& {Makino}(2011){Oshino}, {Funato}, \&
  {Makino}}]{OFM11}
{Oshino} S., {Funato} Y., {Makino} J., 2011, PASJ, 63, 881

\bibitem[{{Portegies Zwart} \& {Boekholt}(2014)}]{PB14}
{Portegies Zwart} S., {Boekholt} T., 2014, ApJ, 785, L3

\bibitem[{{Quinlan} \& {Tremaine}(1992)}]{QT92}
{Quinlan} G.~D., {Tremaine} S., 1992, MNRAS, 259, 505

\bibitem[{{Rein} \& {Spiegel}(2015)}]{RS15}
{Rein} H., {Spiegel} D.~S., 2015, MNRAS, 446, 1424

\bibitem[{{Rein} \& {Tamayo}(2015)}]{RT15}
{Rein} H., {Tamayo} D., 2015, MNRAS, 452, 376

\bibitem[{{Sanz-Serna} \& {Calvo}(1994)}]{SC93}
{Sanz-Serna} J., {Calvo} M., 1994, {Numerical Hamiltonian Problems}, 1st edn.
  Chapman and Hall, London

\bibitem[{{Springel}(2005)}]{S05}
{Springel} V., 2005, MNRAS, 364, 1105

\bibitem[{{Sussman} \& {Wisdom}(2001)}]{SW01}
{Sussman} G.~J., {Wisdom} J., 2001, {Structure and interpretation of classical
  mechanics}. The MIT Press

\bibitem[{{Wisdom}(2006)}]{W06}
{Wisdom} J., 2006, AJ, 131, 2294

\bibitem[{{Wisdom}(2015)}]{W15}
{Wisdom} J., 2015, AJ, 150, 127

\bibitem[{{Wisdom} \& {Hernandez}(2015)}]{WH15}
{Wisdom} J., {Hernandez} D.~M., 2015, MNRAS, 453, 3015

\bibitem[{{Wisdom} \& {Holman}(1991)}]{WH91}
{Wisdom} J., {Holman} M., 1991, AJ, 102, 1528

\bibitem[{{Wisdom}, {Holman} \& {Touma}(1996){Wisdom}, {Holman}, \&
  {Touma}}]{WHT96}
{Wisdom} J., {Holman} M., {Touma} J., 1996, Fields Institute Communications,
  Vol.~10, p.~217, 10, 217

\bibitem[{{Yoshida}(1990)}]{Y90}
{Yoshida} H., 1990, Physics Letters A, 150, 262

\bibitem[{{Yoshida}(1993)}]{Y93}
{Yoshida} H., 1993, Celest. Mech. Dyn. Astron., 56, 27

\bibitem[{{Zhong} \& {Marsden}(1988)}]{ZM88}
{Zhong} G., {Marsden} J.~E., 1988, Physics Letters A, 133, 134

\end{thebibliography}
\end{document}